# Monadic Pavlovian associative learning in a backpropagation-free photonic network

James Y. S. Tan[1†], Zengguang Cheng[1,2†], Johannes Feldmann[1], Xuan Li[1], Nathan Youngblood[1,3], Utku E. Ali[1], C. David Wright[4], Wolfram H. P. Pernice[5,6] and Harish Bhaskaran[1*]

[1]*Department of Materials, University of Oxford, Parks Road, Oxford OX1 3PH, UK.*
[2]*Current address: State Key Laboratory of ASIC and System, School of Microelectronics, Fudan University, Shanghai 200433, China.*
[3]*Current address: Department of Electrical and Computer Engineering, University of Pittsburgh, 3700 O'Hara St., Pittsburgh, PA 15261, USA*
[4]*Department of Engineering, University of Exeter, Exeter EX4 4QF, UK.*
[5]*Institute of Physics, University of Muenster, 48149 Muenster, Germany.*
[6]*Center for Soft Nanoscience, University of Muenster, 48149 Muenster, Germany*

[†]These authors contributed equally to this work.
[*]Corresponding author. Email: harish.bhaskaran@materials.ox.ac.uk

**Abstract**

Over a century ago, Ivan P. Pavlov, in a classic experiment, demonstrated how dogs can learn to associate a ringing bell with food, thereby causing a ring to result in salivation. Today, it is rare to find the use of Pavlovian type associative learning for artificial intelligence (AI) applications even though other learning concepts, in particular backpropagation on artificial neural networks (ANNs) have flourished. However, training using the backpropagation method on 'conventional' ANNs, especially in the form of modern deep neural networks (DNNs), is computationally and energy intensive. Here we experimentally demonstrate a form of backpropagation-free learning using a single (or monadic) associative hardware element. We realize this on an integrated photonic platform using phase-change materials combined with on-chip cascaded directional couplers. We then develop a scaled-up circuit network using our monadic Pavlovian photonic hardware that delivers a distinct machine-learning framework based on single-element associations and, importantly, using backpropagation-free architectures to address general learning tasks. Our approach reduces the computational burden imposed by learning in conventional neural network approaches, thereby increasing speed, whilst also offering higher bandwidth inherent to our photonic implementation.

**Introduction**

The ability to decipher non-trivial patterns in data using computational techniques has led to the development of sophisticated machine intelligence approaches with a plethora of scientifically and technologically important applications [1-4]. Such approaches have predominantly been performed on general-purpose digital electronic processors (i.e., GPUs and CPUs), but this can introduce unwanted and deleterious computational latency and limitations to data throughput. Thus, special-purpose hardware accelerators designed intentionally for use in machine learning applications are an essential development [5-9]. Harnessing the wavelength-multiplexing capabilities of photonics to carry out parallel operations simultaneously in special purpose accelerators can greatly increase the capacity of intelligent information processing [10-13].

Practical associative learning hardware accelerators require a hardware device structure that can associate inputs to a device and current implementations using electronic [14-26], optoelectronic [27], and synthetic biological [28] approaches are limited at the device level. Specifically, the ability to monadically associate at least two inputs together is distinctly absent at a device level. In this paper, we experimentally demonstrate such a single associative learning element, one that exploits the ultra-high bandwidth capabilities of photonics in a readily scalable architecture with the potential to deliver future artificial neural networks with very significantly faster, and lower energy cost training as compared to conventional approaches.

**Associative Monadic Learning Element (AMLE) concept**

In biological systems, a fundamental associative learning process – classical conditioning – can be described using the neural circuit in Fig. 1a [29,30]. For a motor neuron to generate an action potential, it must receive a sensory signal. Pavlov in his experiment showed that salivation in a dog can be "stimulated" by associating the ringing of a Bell with Food [31], i.e. the two sensory stimuli were associated so as to generate an identical response. The process of associating a stimulus $s_2$ from a sensory neuron with the natural stimulus $s_1$ is the process of association, and this is a learning mechanism. Once the association between the two signals is established, the response is triggered when either $s_1$ or $s_2$ is sent to the motor neuron through a synaptic weight ($w_1$ or $w_2$ in Fig. 1a) – the network has now learnt this response. Thus, this simplified neural circuitry has two main roles: one, to converge and associate the two inputs; and secondly, to store memories of these associations *in situ*.

In our optical device, we embed the above functionality in an associative monadic learning element (AMLE) (Fig. 1b). The device has two coupled waveguides and a thin film of phase-change material $Ge_2Sb_2Te_5$ (GST) on the lower waveguide which effectively modulates the coupling between the waveguides. The GST exists in two states (amorphous or crystalline). These two states (and the fractional volumes between the two states) govern the amount of coupling between the waveguides. As shown in Fig 1c, when the material is crystalline, there is no association between the inputs $s_1$ and $s_2$. However, when the two inputs (learning pulses) arrive at the same time, the material has sufficient absorption to amorphize the GST, changing the coupling between the waveguides. As the material amorphizes, inputs $s_1$ and $s_2$ begin to "associate" as shown in Figs 1d. As the number of learning pulses increases, a larger volume of material switches from the crystalline to the amorphous state, until reaching a point when the two inputs result in an output that is nearly indistinguishable – we set this level and term this the learning threshold. Unlike previous non-volatile phase-change material photonic memories which relied primarily on optical amplitudes for their operation [11], we here employ the optical phase difference between inputs $s_1$ and $s_2$ to precisely control the phase state of the GST cell. This enables us to establish the precise extent of association between the inputs. Our phase-change material GST is known to have an ultrafast structural phase transition time (sub-ns amorphization and few-ns crystallization time [32]), high cycling endurance (~$10^{12}$ cycles [33,34]), and long retention time (>10 years at room temperature [34]). A thin capping layer of indium tin oxide (ITO) is deposited on the GST to prevent oxidation, and to help localize optically-induced heat to enable low-power phase switching [11,35].

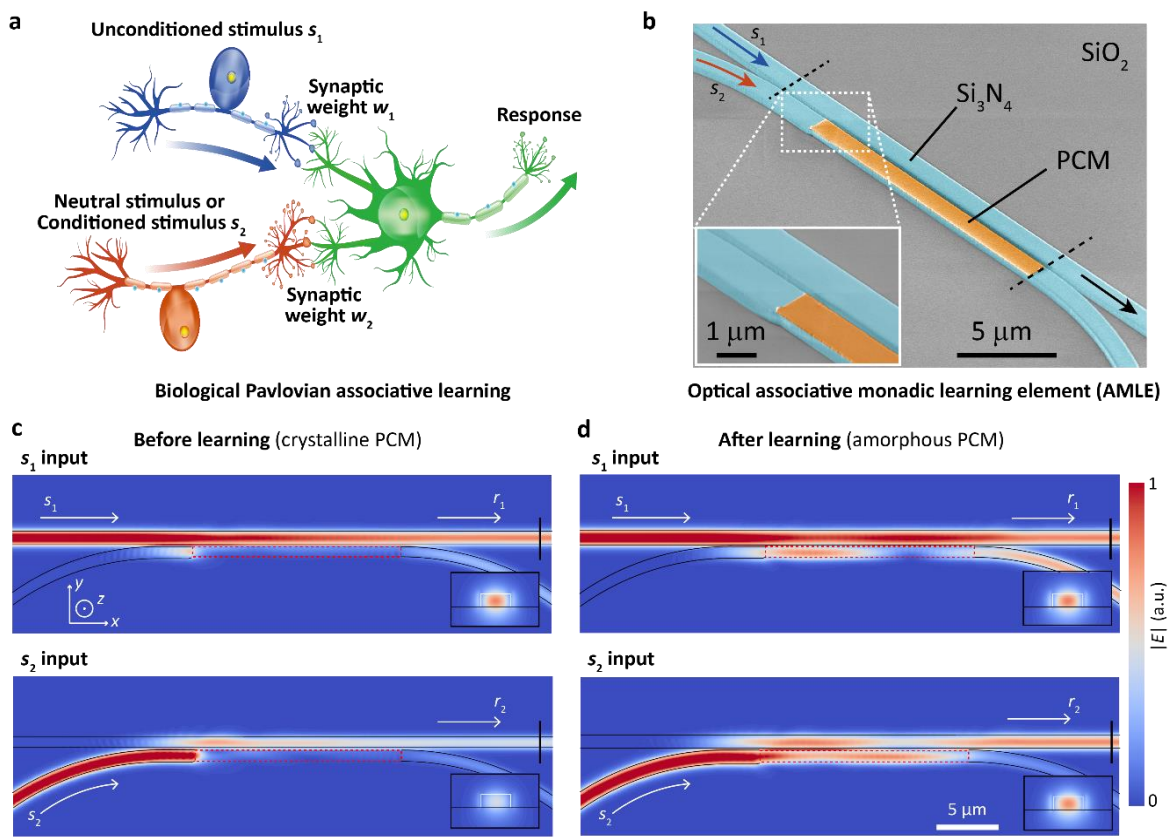

**Fig. 1. Optical associative monadic learning element (AMLE).**
**a**, Simplified illustration of the neural circuitry for associative learning. After the stimulus $s_2$ is associated with stimulus $s_1$, both elicit the same response. **b**, SEM image (false colored) of a fabricated AMLE, consisting of $Si_3N_4$ directional couplers (cyan), $SiO_2$ undercladding (grey) and GST cell (marigold). Inset shows detail of coupling area with GST. **c, d**, Corresponding electric field profiles of AMLE before (**c**) and after (**d**) learning, with $s_1$ and $s_2$ inputs. Inset: Output cross-section optical field at locations denoted by black vertical bars. The color bar is normalized.

The association between inputs $s_1$ and $s_2$ during the learning process occurs only when the two inputs are paired at a specific optical phase delay $\Delta\varphi$. This results in a change in the synaptic weight $\Delta w$ between the stimulus and response signals. In our AMLE, the optical delay $\Delta\varphi$ is introduced by the optical phase difference between the $s_1$ and $s_2$ inputs. We found that the slightest vibration and/or temperature change in the measurement environment can cause the optical phases to vary erratically. We mitigated this by using an on-chip layout which greatly reduced effects of environmental disturbances on the phase control.

The use of directional couplers ensures that the design is applicable over a broad optical wavelength range. Our simulations of the design show the natural response, as outlined in Fig. 1c (prior to associative learning taking place) and Fig. 1d (after the association). Before learning, only $s_1$ leads to high transmission response, whereas $s_2$ does not. After learning, both $s_1$ and $s_2$ produce high transmission response, which indicates that the two inputs are now associated, i.e. the system has "learned" to associate the two inputs such that either triggers the same response.

**Observation of photonic associative learning**

We now experimentally characterize the dynamic response of the AMLE. In order to achieve this, it is necessary to control the input signal combinations ($s_1$ only, $s_2$ only, and both $s_1$ and $s_2$) to the AMLE. We achieve this through the use of wavelength-selective, critically-coupled ring resonators to each of the inputs leading to the AMLE. This allows us to characterize the real-time dynamics of the associative learning process.

The starting point of the AMLE for our experiments is its crystalline state. We probe the output transmission $r_1$ and $r_2$ of the AMLE in real-time using wavelengths $\lambda_1$ and $\lambda_2$. As shown in Fig. 2a, the transmission readouts $r_1$ and $r_2$ remained the same for single input pulses at 1.45 nJ (pulse widths $\tau$ = 100 ns) as expected (for events 1 to 4). However, when inputs $s_1$ and $s_2$ were sent simultaneously at pump wavelength $\lambda_0$ with a fixed phase delay of $\pi/2$ at 0.66 nJ ($\tau$ = 100 ns) each in event 5, the transmission change $\Delta r_1$ and $\Delta r_2$ for the $s_1$ and $s_2$ probe readouts are ~ −4% and ~ +4% respectively. As the input pump pulse power was increased from 0.87 nJ ($\tau$ = 100 ns) to 1.45 nJ ($\tau$ = 100 ns) in events 6 to 8, the probe readouts changed by approximately −7% and +7% respectively, both of which are well above our output transmission threshold of $r_{th}$ ~ 5%. Details of optical pulses are provided in Appendix. Effectively, these experiments show that the two inputs can be "taught" to associate with each other such that either triggers the response, i.e. the associations are learned.

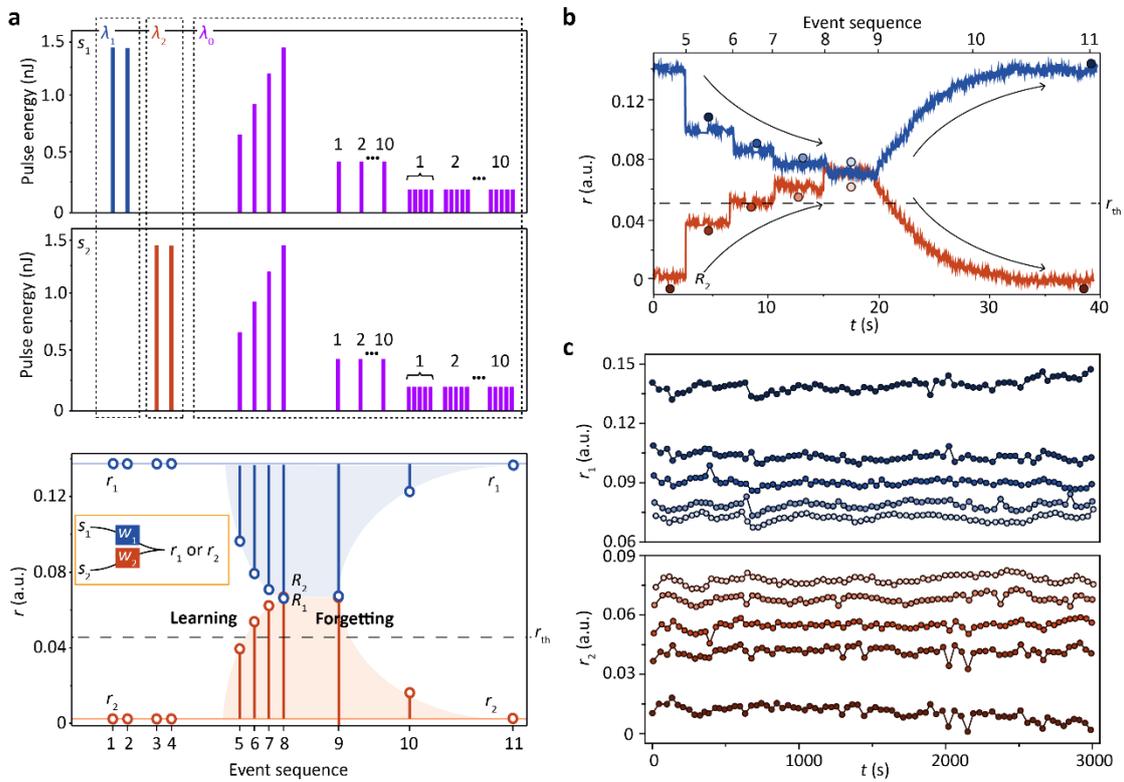

**Figure 2. Photonic Pavlovian learning process.**
**a**, Input-output relation of AMLE. $s_1$ and $s_2$ inputs denote 'Food' and 'Bell' inputs; the corresponding output transmission $r$ (bottom chart) represents the transmission response $r_1$ and $r_2$. Blue, red, and purple bars of the top and middle charts denote $s_1$, $s_2$ and both $s_1$-$s_2$ input incidences ($\lambda_1$, $\lambda_2$ and $\lambda_0$ are the wavelengths used to selectively address each one). Bottom chart inset: Simplified diagram of AMLE. **b**, Corresponding real-time measurement of output probe transmission $r$ of a single cycle learning and forgetting processes in (**a**). **c**, Repeatability of the processes on AMLE over 80 cycles. The levels are denoted by filled circles of different colors that correspond to (**b**).

We then show in Fig. 2a (bottom chart) that these learned associations can also be reversed. A set of pulses at 0.43 nJ ($\tau$ = 100 ns) in event 9, followed by 0.19 nJ pulses ($\tau$ = 100 ns) in event 10 resulted in the 'forgetting' process, where the readouts $r$ reverted to the baselines ($r_1$ ~ 0.14 for $s_1$ input probe and $r_2$ ~ 0 for $s_2$ input probe). For our measurements, readout above a threshold $r_{th}$ ~ 0.05 is designated as the learned state.

Fig. 2b shows a single cycle of the real-time output readout of associative learning in events 5 to 8 and the forgetting process in events 9 to 11 of Fig. 2a. To test the repeatability of our associative learning and forgetting processes, we subjected the AMLE through 80 learning cycles, examined over a period of 50 minutes. After 80 cycles (Fig. 2c), the individual learning weights were clearly identifiable with standard deviation of ±0.69% in readout transmission.

**Associative network for supervised learning**

Up to this point, we have observed associative learning and characterized the workings of our AMLE via single device photonic measurements. The general concepts of Pavlovian associative learning (Fig. 3a) and supervised learning (a class of machine learning; Fig. 3b) are in essence comparable – both involving the pairing of input (*IN*) with the correct output (*Teacher*) to supervise the learning process. However, the mapping of the input to the desired output in conventional supervised learning architectures is relatively complicated as well as time and energy consuming – the *Teacher* signal propagates backward layer-by-layer to collectively adjust the network weights such that the actual output *R* better resembles the desired output (*Teacher*) after each learning iteration. The learning process for our AMLEs is by comparison far more straightforward (and faster and more energy efficient), and to elucidate this we consider a scaled-up network of AMLE devices, which we illustrate schematically in Fig. 3c.

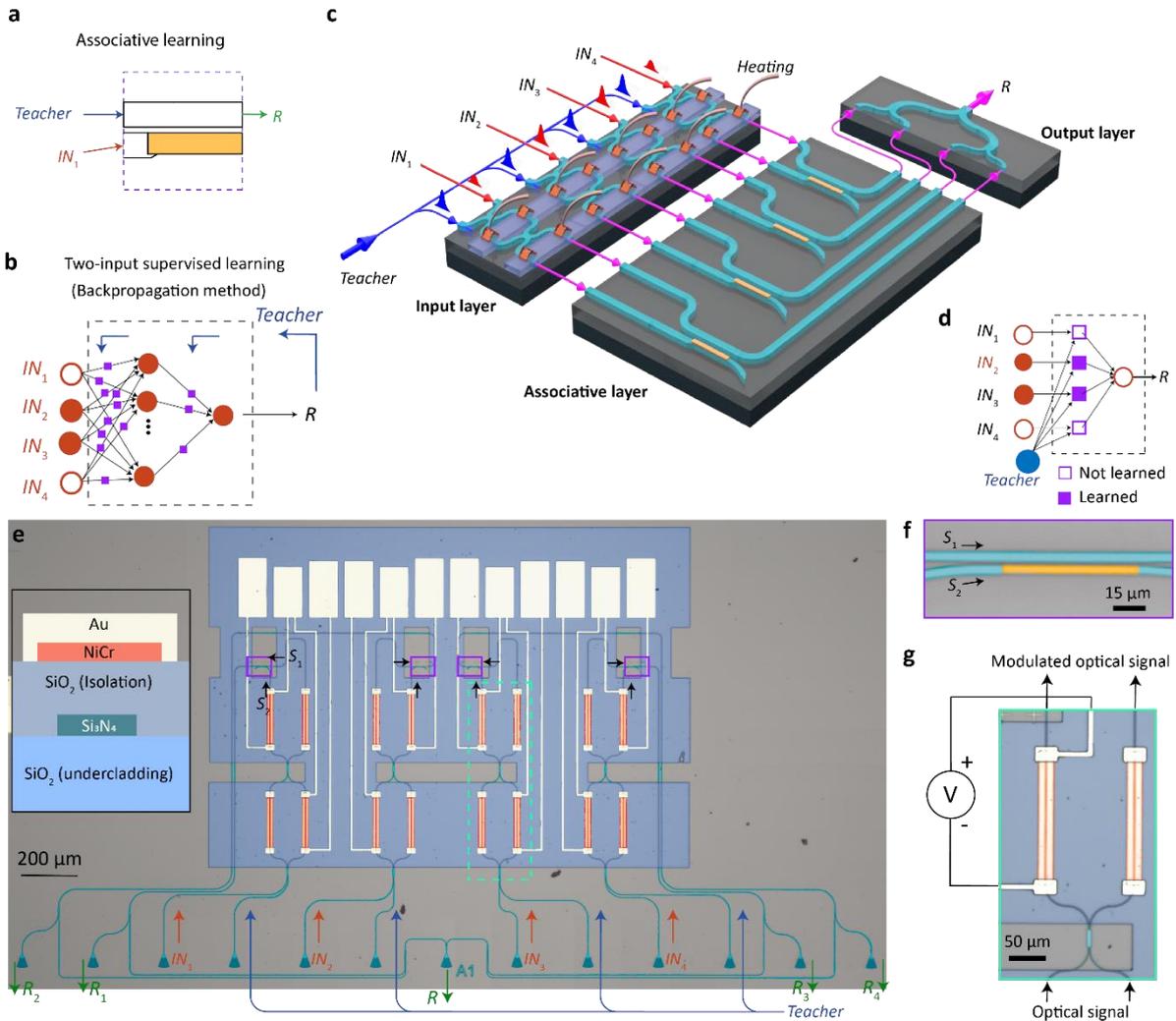

**Figure 3. Supervised Pavlovian associative learning.**

**a**, Pavlovian learning involves pairing the inputs (*IN*) with the correct outputs (*Teacher*) to supervise the learning process. **b**, Current conventional supervised learning networks use backpropagation. The network diagram depicts the network for supervised learning. **c**, Optical on-chip hardware diagram of supervised learning with two inputs using AMLE. Input signals $IN_1$ to $IN_4$ fed into the system during the learning process to be supervised by the *Teacher* input signal. $IN_1$ to $IN_4$ and *Teacher* inputs leading to the AMLE are controlled by thermo-optic (with NiCr heaters) MZMs. **d**, Network representation of the hardware diagram in (**c**). **e**, Optical micrograph of supervised learning network in (**c**), which consists of four AMLEs (boxed, purple). The arrows show the optical input-output connections coupled to the on-chip network via grating couplers. **f, g**, Optical micrographs of a single AMLE and heaters respectively correspond to those in (**e**).

To implement the system of Fig. 3c, we use cascaded Mach-Zender Modulators (MZMs) to provide a reliable means to split both the *Input* and *Teacher* signals equally with stable optical phases (obtained via the use of integrated NiCr thermo-optic heaters) and feed them to the inputs of the AMLEs. The MZMs also allow for the use of wavelength multiplexing to feed multiple signals to the inputs of multiple AMLEs, before these signals are paired with *Teacher* signals at the associative layer to cumulatively produce and output transmission response. A more detailed illustration of the resulting integrated AMLE network chip is thus as shown in Fig. 3e. The system consists of the associative layer (purple bordered boxes) between an input layer (*Input* data and *Teacher* denoted using red and blue arrows respectively) and an output layer (response *R*; green arrow). Optical micrographs with an enlarged view of the AMLE and thermo-optic heaters are shown in Fig. 3f, and Fig. 3g respectively.

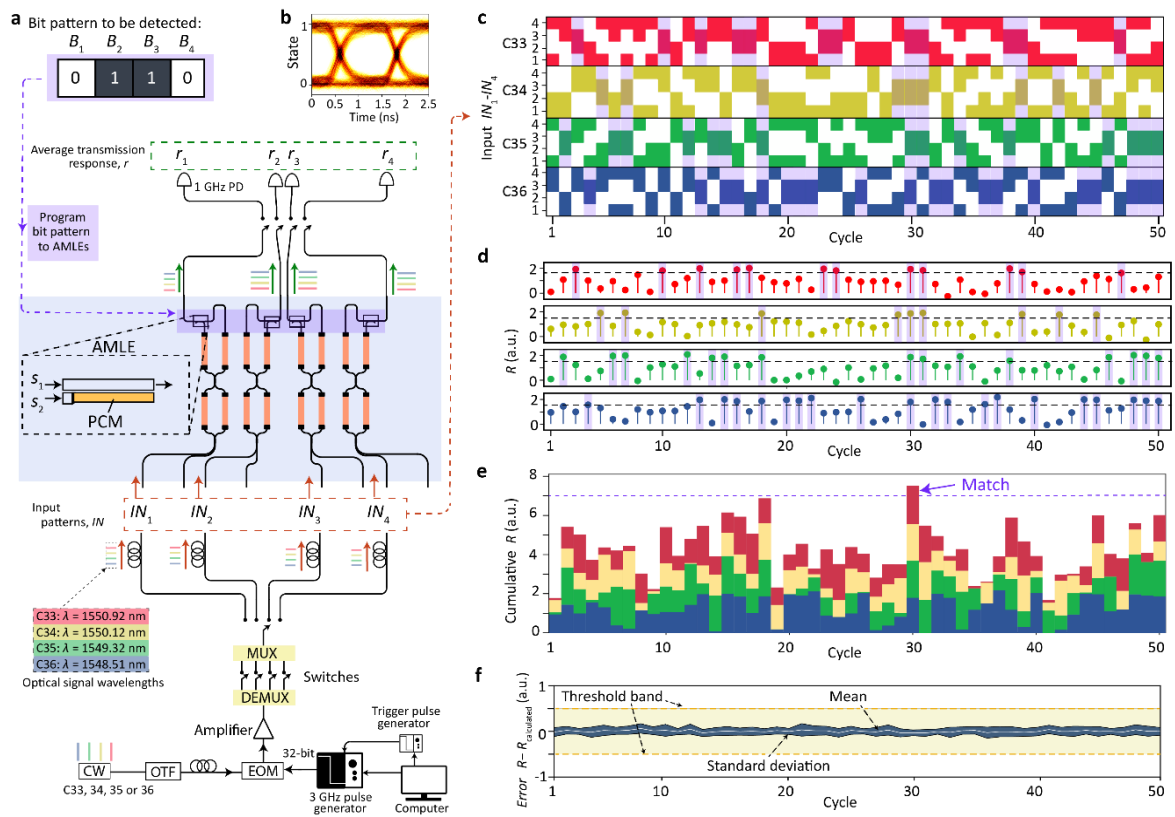

**Figure 4. Pattern recognition with AMLE.**

**a**, Measurement setup for pattern recognition. Bit patterns to be detected '0110' are associated into the AMLE via a learning step. 1 GHz optical detection pulses defined by computer-generated pseudo-random bits at wavelengths from C33 to C36 (1548.51 nm to 1550.92 nm) are sent through AMLE devices and detected by photodiodes as the signals are temporally-multiplexed. Measured average output transmission *R* cumulatively summed to determine if pattern is recognized. **b**, Eye-diagram obtained from non-return-to-zero (NRZ) pseudo-random binary sequence (PRBS) pulses modulated at

1 GHz shows clear distinction between states '0' and '1'. **c**, Inputs $IN_1$ to $IN_4$ sent at the four different wavelengths i.e., C33 to C36 (input = '1' represented in red, yellow, green and blue respectively). Purple shades represent '0110' pattern sent to the devices. **d**. $R$ for the different wavelengths (C33 to C36). Purple shades represent '0110' pattern detected. **e**, Cumulative $R$ obtained for the respective wavelengths and detection cycles. Values above cumulative transmission threshold $R = 7$ (the dashed black line) denote all-matching response (i.e., in cycle 30). **f**, Calculated error $R - R_{calculated}$ (difference between measured average transmission response and expected response) shows mean error is negligibly low with standard deviation well within the ±0.5 threshold band.

We use the AMLE network chip of Fig. 3e to carry out a rapid pattern recognition task to verify that once the associations are formed by the AMLEs, simultaneous parallel recognition can be achieved. The GST cells are initially set to their crystalline states (details in Appendix). We then program the bit pattern '0110' to the AMLEs by respectively sending the pump *Teacher* signal $T = 1.47$ nJ ($\tau = 100$ ns) and pump *Input* signals $IN_1 = 0$, $IN_2 = 1.47$ nJ ($\tau = 100$ ns), $IN_3 = 1.47$ nJ ($\tau = 100$ ns) and $IN_4 = 0$ simultaneously – essentially associating these patterns within the AMLE network. Pattern recognition is then carried out by sending a series of randomized optical binary signals modulated at 1 GHz speed to the AMLEs (with 0 mW for '0' and 0.7 mW for '1'), using different wavelengths (in the range 1548.51 nm to 1550.92 nm, C33 to C36 in the telecoms band) for each of the individual AMLE inputs – see Fig. 4a. The output from each AMLE device in the network passes to four photodiodes to generate the system outputs (or responses), $R$. The output transmission response of a single AMLE $r_{ij}$ measured at wavelength $\lambda_j$ from the photodetectors is then summed to obtain the cumulative transmission response $R_j = \Sigma_i R_{ij}$ for the specific wavelength, where $R_{ij}$ is the measured output transmission of the $i^{th}$ AMLE and $j^{th}$ wavelength after normalizing (between 0 and 1 range) using $R_{ij} = (r_{ij} - r_{base})/r_{base}$, i.e. the average value of two consecutive output transmission values using a trigger pulse (with $r_{base}$ being the baseline of the optical transmission. The average value is taken to minimize the effects of electrical or optical noise. After recording the output pulses from the experiments at the four optical wavelengths, we systematically retrieve the cycle(s) at which the cumulative response $R_c = \Sigma_{ij} R_{ij}$ is above a pre-determined learning threshold. Fig. 4c shows the inputs sent at the four different wavelengths C33 to C36, while the corresponding output response of the AMLEs is shown in Fig. 4d. Fig. 4e shows the cumulative $R_c$ for all four wavelengths, which exceeds the learning threshold only at cycle 30; this enables us to pinpoint the output bit combinations at which input signals from all four wavelengths match the bit pattern '0110' to be at cycle 30. Fig. 4f shows the performance of the AMLEs for detection predictions. We measured the prediction error $Error = R - R_{calculated}$ from the difference between output transmission $R = \Sigma_i R_{ij}$ and the calculated transmission $R_{calculated}$. The error is within the ±0.5 threshold bands.

It is important to point out that the detection speed in the experiments shown in Fig. 4 is limited only by the number of pulses to average (in our case, two) and the signal modulation and detection speed of the modulators and photodetectors. In principle, this detection speed can be significantly improved by increasing signal to noise ratio (by using on-chip photodetectors instead of external photodetectors connected via grating couplers in our case) and increasing the input signal modulation speed while maintaining detected signal integrity. Although we demonstrated detection specifically for the pattern '0110', this hardware system can be used to detect all other possible (and more) patterns by additionally representing on other AMLEs another separate set of bits toggled to the bit pattern to be detected, and then additionally sending toggled sets of input bits to these additional AMLEs.

Now, we demonstrate how AMLE-based hardware can achieve generalization on an image recognition task using associative learning, based on the network architecture shown in Fig. 5a. The network is similar to those in Fig. 3d, which consists of three main network layers (i.e., input layer, associative layer and output layer) [36]. During the training process, images to be trained are first pixelated to 15×13 input pixel data (195 pixels altogether). These data $IN_1$-$IN_{195}$ are then sent to the associative layer, whilst simultaneously the *Teacher* signals are likewise preprocessed and sent to the associative layer. Notably, images fed into $IN_1$-$IN_{195}$ and *Teacher* signals are exchangeable in the training process. The change in the learning states of AMLEs is determined from their transmission readouts, when preprocessed *Input* data of maximum amplitude (blank white image) and the same dimensions as the training pixel data is fed through the layers. The transmission of these individual AMLEs is summed up at the output layer and rearranged to form a 15×13-pixel model representation. The complete computation is obtained by cumulatively adding the model representation from each training pair. Pixel-by-pixel comparison between the model (which generalizes the training images) and measurement is then made to determine if the testing images are of the image to be detected.

The photonic implementation of the associative learning network architecture for the image recognition is shown in Fig. 5b. Here, the GST elements on the AMLEs are first initialized to the crystalline state before the signals are sent to the on-chip photonic structure. During training, depending on the optical input signal and the input pump power sent to the AMLEs, the state of the GST cells on AMLEs either remain in crystalline, or structurally switch to amorphous state. This results in a change in optical probe output response of the AMLEs. In our experiment, because we only have 4 AMLEs, we raster the 195 pairs of input pixels across the 4 AMLEs.

We examine the cat image classification capabilities of our associative learning network using the 'Dogs vs Cats' dataset from fast.ai [37] (single dataset collected from CIFAR10, CIFAR 100 [38], Caltech 101 [39], Oxford-IIIT Pet [40] and Imagewoof [41]). The cat images that we used for the training process are shown in Fig. 5c. After the training process using these images, we obtain the model representation of a cat shown in Fig. 5d by feeding in a 15×13-pixel blank white test input of maximum probe input power magnitude (1.3 mW).

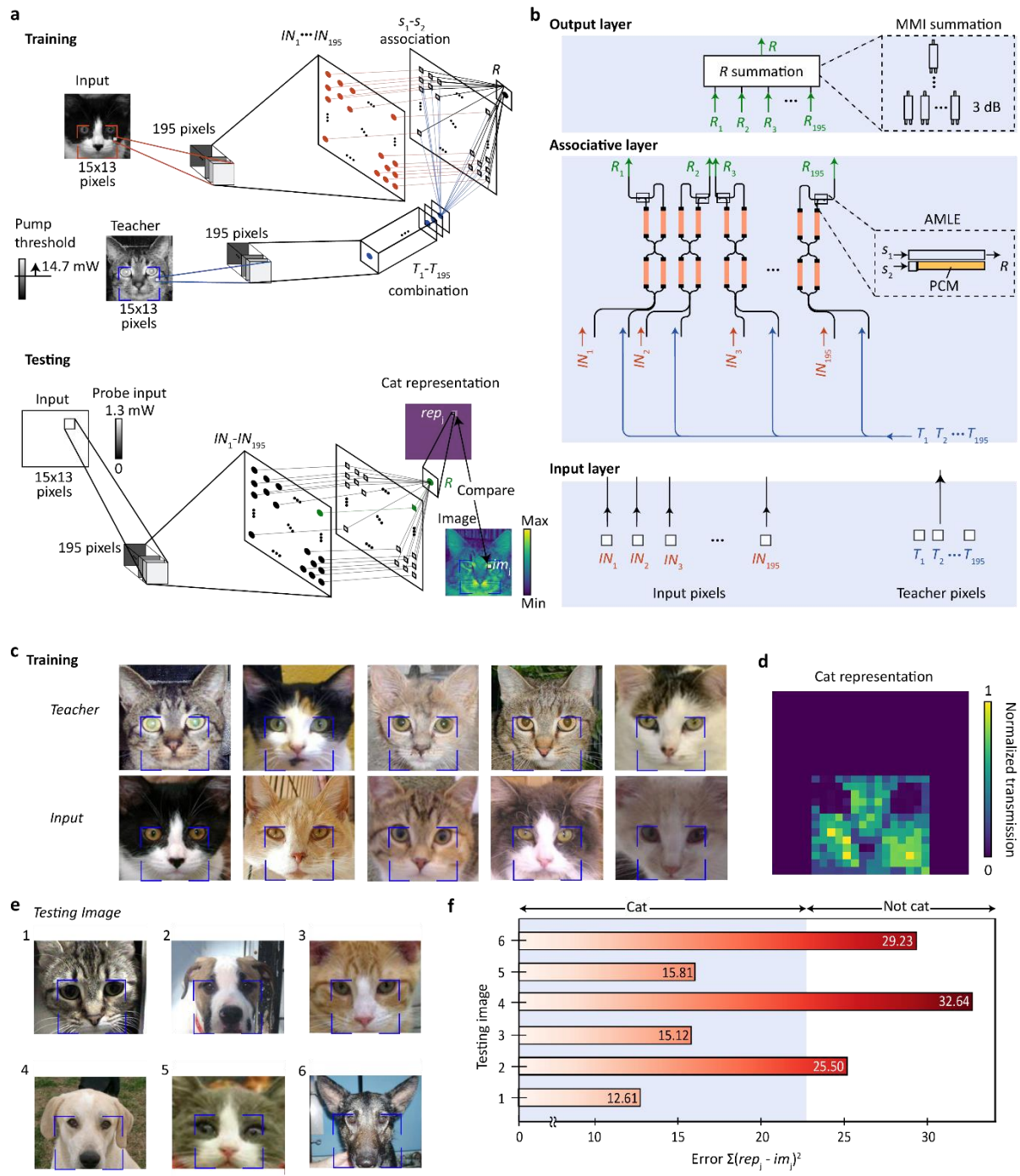

**Figure 5. Scaling architecture for image recognition using associative learning network.**

**a**, The general associative neural network composed of an input layer ($IN_1$-$IN_{195}$, $T_1$-$T_{195}$), associative layer (stimuli $s_1$-$s_2$ association) and output layer (transmission $R$). The input signal is the pattern (pixels from image) to be classified, and the external teacher provides the desired set of output values chosen to represent the class of patterns being learned. During training, the input layer ($IN_1$-$IN_{195}$, $T_1$-$T_{195}$) is fed into the associative layer. Associative layer consists of AMLEs with states that are modified when both the input and external teacher signals are paired together. A model representation of the trained images is then obtained by sending a preprocessed input signal of blank white image to propagate through the layers. **b**, Optical on-chip implementation of the input layer, associative layer and output layer respectively. The associative layer, which consists of thermo-optic NiCr heaters, distributes the combination of the input and external teacher signals (from the input layer) as $s_1$ and $s_2$ inputs to the respective AMLEs. The output layer consists of a summation unit to sum up the output response from the AMLEs to form the net output. Conventional 2×1 combiner has 3 dB loss each. **c**, Training the associative learning network to identify cat images, with bounding box corners to indicate region of interest. **d**, After five training iterations, the network learns the model representation of cat from the output response $R$. **e**, Images used to test if the network can correctly classify pictures as cat and non-cat. **f**, The network successfully recognized cat and non-cat images based on the squared Euclidean distance $\Sigma\,(rep_j - im_j)^2$ measured over the pixelated testing images for each of the 15×13 pixels.

Thus far, we demonstrated searching for patterns (in the form of pixel amplitudes) from the 15×13-pixel image sent to the network of monadic AMLEs in a single step. After each training iteration with a model image, the network's ability to distinguish the appearance of a cat improves. The feature subtleties can then be captured from the model representation, giving us a valuable means to distinguish a cat from other objects. To test the model representation, we use the testing images shown in Fig. 5e to compare them ($im$) with the representation of the cat $rep$ and measure the error function with respect to the pixel $j$, given by $tanh\,((rep_j - im_j)^2\,e)$ for every pixel $j$. We set the threshold $min(Error) + (max(Error) - min(Error))/2$, which is 22.625 for our case, to determine the *Testing Images* that resemble the model representation in Fig. 5d. Our results, summarized in Fig. 5f, reveal that *Testing Image* 1, 3 and 5 resemble the cat representation. With the error ($Error$ = squared Euclidean distance $\Sigma\,(rep_j - im_j)^2$) of Image 2, 4, 6 above the threshold, the network predicts that these images are not of a cat. The associative learning network thus accurately classifies images of cats from the model representation obtained from the training iterations. In contrast to conventional numeric-based AI approaches, our symbolic associative learning system interprets from the 'cat representation' results to logically conjure up the proposition that an image must have the shown clear, distinctive, and comprehensible physical attributes to constitute a cat image. Our image recognition example adopts symbolic AI while conventional image recognition uses connectionist AI. Our approach is typically simpler and faster, but comes at the expense of not having the ability to acquire deep features – an ability which may not be needed for many less complex machine learning tasks.

**Evaluation metrics for associative learning devices**

Our application-specific system offers convergence in one training step due to its straightforward approach of detecting similarities without having to randomize network weights and perform back-propagation as in conventional neural networks.

**Table 1.** Comparison of active volume and learning energy in associative learning devices

| Type | Active volume (µm$^3$) | Min. learning energy (nJ) | Ref. |
|---|---|---|---|
| **Electronic** | | | |
| ▪ Memresistive | | | |
| i. Chalcogenide | 0.12 – 10.5 | $4.7 \times 10^4$ | [14] |
| | 8 | 2.63 | [15] |
| ii. Manganite | ~ 0.1 | $1.35 \times 10^3$ | [16] |
| | $1.25 \times 10^{10}$ | $1.02 \times 10^5$ | [17] |
| iv. Nickelate | $4.7 \times 10^3$ | $7.20 \times 10^5$ | [18] |
| | $4.8 \times 10^4$ | $2.04 \times 10^5$ | [19] |
| v. Metal oxide | ~10 – 900 | ~ $10^3 - 10^5$ | [20,21] |
| vi. Organic | ~ 0.1 – 0.5 | ~ $10^3 - 10^4$ | [22,23] |
| ▪ Electrochemical | $6 \times 10^3$ | $6 \times 10^4$ | [24] |
| | $9.6 \times 10^5$ | 125 | [25] |
| ▪ Memcapacitive | 26.9 | ~ 30 | [26] |
| **Optoelectronic** | $1.62 \times 10^3$ | $2.1 \times 10^3$ | [27] |
| **Optical AMLE** | 0.12 | 1.8 | (**this work**) |

We identify relevant device-level evaluation metrics by contextualizing the AMLE with a typical machine learning data load, which requires data to be transferred back and forth from the data source to be run using cloud computing and/or supercomputers. For a more energy-efficient locally-run neural network, it is important to shrink the network for greater portability and reduce the energy consumption of the learning process. Table 1 summarizes the minimum active device volume (in our case, volume of GST and the waveguide below) and learning energy of other associative learning devices [14-27]. The electronic and optoelectronic associative learning devices range from ~ 0.1 - $10^{10}$ µm$^3$ in active volume and consume ~ 2.63 - $10^5$ nJ of energy per learning event [14-27]. In comparison, the all-optical AMLE in our work fares favorably relative to these devices in terms of dimensions and very favorably in terms of energy usage, with a low active volume at 0.12 µm$^3$ and minimum learning energy at only 1.8 nJ. The overall size of the single-element device is of dimensions 2 µm × 17 µm × 0.33 µm. Despite the relatively low active volume (of GST) in our case, compared to electronic hardware, our devices are still larger; however, the speed and multiplexing of photonic devices can lead to higher computational density. Experimentally, we have demonstrated 118 TOPS/mm² which was limited by the available setup in the lab. Scaling the device to more wavelength channels and modulating at higher speed potentially leads to significant gains in the compute density. Considering for example modulation at 50 GHz [42] and using 16 wavelengths, the device would deliver approximately $2.5 \times 10^4$ TOPS/mm². The overall net energy usage of the learning system is directly related to the number of iterations required to train a learning system. The energy-speed tradeoff – particularly evident in CMOS devices [43] – is another important evaluation metric that requires further investigation and research. Our device learns in ~ 100 ns, compared to ~ ms in a previous associative learning device [15].

**Conclusion**

Our results show the first demonstration of an associative monadic learning element (AMLE) implemented on a photonic platform. We provide a supervised learning framework that facilitates the transition from a monadic Pavlovian single *Input-Teacher* association on an AMLE to any arbitrary *n Input-Teacher* associations, thus enabling backpropagation-free, single-layer weight artificial neural network architectures.

We have elucidated the inner workings of the network building block, that can spatiotemporally correlate two initially distinct inputs ($s_1$ and $s_2$) to a same output when both the inputs are simultaneously applied at a predetermined $\Delta\varphi$ optical delay. Given that light signals inherently do not interfere at different wavelengths in linear media (including the AMLE), such input-input association can handle associations of multiple data streams consisting of different wavelengths over a single element, as we have experimentally demonstrated. Our photonic platform allows for wavelength-multiplexing, which is inherently suited to the highly parallel nature of machine learning. We anticipate further improvements in other relevant metrics (e.g., overall device volume and learning energy) on different materials platform and with other optimization methods.

More generally and interestingly, our work can extend the one-way learning ($s_2$ becomes associated to $s_1$) to a customizable form of learning, for example mutual / two-way learning ($s_2$ becomes associated to $s_1$ while $s_1$ is associated to $s_2$). This customizable feature when combined with demonstrations of deterministic weights using identical, fixed energy and fixed-duration pulses [35], will provide unprecedented design flexibility for a wide range of machine learning applications. As to whether the nonlinear scaling with the number of inputs for nonlinear classification problems is an inherent attribute in associative learning is still an open question. However, as shown in Fig. 4 and Fig. 5, practical applications such as pattern recognition and image recognition can be readily demonstrated with associative learning network that scales linearly with the number of inputs. The compact single-element implementation in our work will allow the use of the AMLE as a building block in machine learning/statistical inference in general, thus potentially opening up new avenues of research in machine learning algorithms and architectures.

**APPENDIX: METHODS**

**1. FDTD simulation**

Three-dimensional finite-difference time-domain (FDTD) simulations were performed using the FDTD Solutions software from Lumerical Inc. A fundamental quasi-transverse electric (TE; magnetic field component $H_z$ dominant) optical mode source input of 1 V/m at 1.58 µm wavelength is let incident onto the AMLE waveguide input. The simulation plots in Fig. 1,c,d show optical field $|E|$ profile taken at the central cross section in the *x-y* plane and *y-z* plane (axis depicted in Fig. 1c) of the AMLE structure. Our numerical simulations summarized in Fig. S1, g-h indicate that the output transmission at 1.55 µm wavelength is within the same range as those at 1.58 µm wavelength.

**2. Device fabrication and characterization**

The AMLE is fabricated on a $Si_3N_4/SiO_2$ platform. Electron beam lithography (JEOL 5500FS, JEOL Ltd.) is used at 50kV to define the $Si_3N_4$ structure on the Ma-N 2403 negative-tone resist-coated substrate. After the development process, reactive ion etching (PlasmaPro 80, Oxford Instruments) is performed in $CHF_3/O_2/Ar$ to etch down 330 nm of $Si_3N_4$. Electron beam lithography is then implemented on a poly(methyl methacrylate) (PMMA) positive resist-coated substrate to open a window for the GST cell. This is followed by sputter-deposition (Nordiko RF Sputter Tool) of 10-nm GST/10-nm ITO on the substrate. For the heater-based layout, windows to deposit $SiO_2$ isolation layer are opened on photoresist S1813, exposed using mask aligner (SUSS MicroTec SE). An NiCr layer (for heaters) are deposited on the SiO$_2$ isolation layer using the sputtering tool. Gold layer (for on-chip electrical pads) are deposited on NiCr layer using thermal evaporation (Edwards 306 Vacuum Coater/Deposition systems). The AMLE characterization process is performed using high resolution emission gun SEM (Hitachi S-4300 SEM system- Ibaraki, Japan) with low accelerating voltage (1 to 3kV) at working distance of ~13mm, and using optical microscope (Eclipse LV100ND, Nikon).

### 3. Optical measurement

The experiment setup to measure AMLE on ring-based layout (Layout 1; for Fig. 2) builds upon previously described probe-pump configuration [11]. To probe AMLE transmission, two low-power continuous-wave (CW) probe diode lasers (N7711A, Keysight Tech.) are used as probe lasers. The signals coupled from the layout are filtered by optical tunable band-pass filters (OTF; OTF-320, Santec Corp.) and detected by photodetectors (PD; 2011-FC, Newport Spectra-Physics Ltd.). To induce learning on the AMLE, optical pump pulses are sent to the AMLE. A CW diode laser (TSL-550, Santec Corp.) is used for the pump signal. The pulse shape of the optical signal is defined by electro-optic modulator (EOM; 2623NA, Lucent) based on the electrical pulse shape generated by the arbitrary function generator (AFG; AFG 3102C, Tektronix). The optical pump pulses are then amplified by a low-noise erbium-doped fiber amplifier (EDFA; AEDFA-CL-23, Amonics) and sent to the AMLE.

In the stabilization step carried out prior to the experiment, a set of amorphizing pulses are sent to the AMLE, followed by a set of crystallizing pulses. These sets of pulses are exactly the same as the pulses applied during the 'associative learning' and 'forgetting' process shown in Fig. 2a. Here, the set of amorphizing pulses are the consecutive 100 ns-wide pulses at 0.66 nJ, 0.87 nJ, 1.26 nJ and 1.45 nJ, while the set of crystallizing pulses are the 100 ns-wide pulse at 0.43 nJ for ten times, followed by five 0.19 nJ 100 ns-wide pulses at 1 MHz repetition rate for ten times. These forgetting pulses introduce sustained heating at temperatures above crystallization temperature and below melting temperature to crystallize the PCM.

Our pattern and image recognition experiments in Fig. 4 and Fig. 5 are carried out by measuring the associative hardware network shown in Fig. 3e. These experiments are based on the experiment setup to measure a single AMLE device on heater-based layout (Layout 2).

The pattern recognition experiment in Fig. 4 consists of two steps: pattern programming and pattern detection. In the first step, we program the AMLE with bit patterns by sending pump pulses to the AMLEs in the associative hardware network. In the second step, we send probe pulses at high speed (at 1 GHz modulation rate) and measure the AMLE output transmission responses to determine if the patterns sent matches the pattern programmed into the AMLEs. The experiment setup to program a set of patterns to the AMLEs is shown in Fig. S4, a and b. The experiment setup for the detection step is shown in Fig. 4a. In the setup, the AFG (AFG 3102C, Tektronix) specifically picks the continuous 32-bit 1 GHz pulses from the 3 GHz pulse generator (HP 8133A, Hewlett Packard). The EDFA (AEDFA-CL-23, Amonics Ltd.) amplifies the pulse from EOM (2623 NA, Lucent Tech.). The AMLE output pulses coupled from the layout are detected by the 1 GHz fiber optic photodetector (1611FC-AC, Newport).

The image recognition experiment in Fig. 5 consists of two steps: training and testing. As in other face recognition methods, the datasets are preprocessed to filter out irrelevant datasets to ensure that they have reliably sufficient facial landmarks; before the associative learning method engages in the image recognition process. In the setup, the input signals $IN_1$-$IN_{195}$ are represented as optical signals wavelength-multiplexed from the spectrally filtered (SuperK Split, NKT Photonics) laser (WhiteLase Micro, NKT Photonics). Another set of separate supervisory *Teacher* signals $T_1$-$T_{195}$ from CW laser (TSL-550, Santec Corp.) are distributed to each inputs. These input pairs are rastered across the 4 hardware AMLEs shown in **Fig. 3e**. The combined signals are selectively routed to either probe line or pump line amplified by EDFA (FA-15, Pritel and FA-33-IO, Pritel), filtered by OTF (OTF-320, Santec Corp.), and channeled to the layout. At the output end of the layout, the readouts are filtered by OTF (OTF-930, Santec Corp.) and detected by PD (2011-FC, Newport Spectra-Physics Ltd.). During pumping, the optical attenuator (V1550PA, Thorlabs Inc.) connected prior to the OTF is activated to filter out the pump signals from the PD. To modulate optical inputs to the AMLEs in the associative hardware network, voltage biasing to the on-chip NiCr waveguide heaters are applied. The testing process is carried out by comparing pixel-by-pixel the testing image with the net cat representation.

**Acknowledgements:** The authors acknowledge discussions with and the experimental work of Andrew Katumba, Nikolaos Farmakidis, Eugene J.H. Soh, Wen Zhou, and Hazim Hanif. All authors acknowledge funding from the European Union's Horizon 2020 Research and Innovation Programme. This research was also supported via the Engineering and Physical Sciences Research Council. Z.C. acknowledges support from National Key Research and Development Program of China, NNSF China, Science and Technology Commission of Shanghai Municipality, and The Young Scientist Project of MOE Innovation Platform. **Data and materials availability:** Additional data related to this paper may be requested from the authors.

**Disclosure:** The authors declare no conflicts of interest.